\documentclass{ws-procs9x6}
\newcommand{\beq}{\begin{equation}}
\newcommand{\eeq}{\end{equation}}
\newcommand{\bea}{\begin{eqnarray}}
\newcommand{\eea}{\end{eqnarray}}
\def\simlt{\buildrel < \over {_{\sim}}}
\def\simgt{\buildrel > \over {_{\sim}}}
\begin{document}

\title{String Inspired Neutrino Mass Models}

\author{S. F. King}

\address{Department of Physics and Astronomy, \\ 
        University of Southampton, Southampton SO17 1BJ, U.K.\\
E-mail: king@soton.ac.uk}


\maketitle

\abstracts{I discuss a class of hierarchical neutrino mass models based on 
the see-saw mechanism with single right-handed neutrino
dominance. I apply this mechanism to 
a string inspired SUSY Pati-Salam model and indicate how it
may emerge from intersecting D-branes.}

\section{Introduction}

Recent SNO \cite{Jelley} results when combined
with other solar neutrino data especially that of Super-Kamiokande
strongly favour the large mixing angle (LMA) MSW solar solution 
\cite{Smirnov} with three active light neutrino states, and
$\theta_{12} \approx \pi/6$, 
$\Delta m_{21}^2\approx 5\times 10^{-5}{\rm eV}^2$
\cite{Smirnov}.
The atmospheric neutrino data is consistent with
maximal $\nu_{\mu}- \nu_{\tau}$ neutrino mixing
$\theta_{23} \approx \pi/4$
with $|\Delta m_{32}^2|\approx 2.5\times 10^{-3}{\rm eV}^2$
and the sign of $\Delta m_{32}^2$ undetermined. 
The CHOOZ experiment limits $\theta_{13} \simlt 0.2$
over the favoured atmospheric range \cite{Smirnov}.

The leading order Majorana mass matrix consistent with a neutrino
mass hierarchy and bi-large mixing angles is 
\beq
m_{LL} \sim \left(
\begin{array}{ccc}
0 & 0 & 0 \\
0 & 1 & 1\\
0 & 1 & 1 \\
\end{array}
\right)\frac{m}{2}
\label{hier}
\eeq
However since  
$m_3 \sim \sqrt{|\Delta m_{32}^2|}\sim 5.10^{-2}$ eV and 
$m_2 \sim \sqrt{|\Delta m_{21}^2|}\sim 7.10^{-3}$ eV, 
compared to the natural expectation $m_2 \sim m_3$,
it remains a puzzle why there should be any hierarchy at all.
The question may be phrased
in technical terms as one of understanding why 
the sub-determinant of the mass matrix is small:
\beq
det \left(
\begin{array}{cc}
m_{22} & m_{23}\\
m_{23} & m_{33} \\
\end{array}
\right)\ll m^2.
\label{det}
\eeq

\section{The see-saw mechanism with single right-handed neutrino dominance}

The simplest way to generate neutrino masses
from a renormalisable theory is to introduce right-handed neutrinos.
Since the right-handed neutrinos are electroweak singlets
the Majorana masses of the right-handed neutrinos $M_{RR}$
may be orders of magnitude larger than the Dirac
masses $m_{LR}$. In the approximation that $M_{RR}\gg m_{LR}$ 
one obtains effective Majorana masses
\beq
m_{LL}=m_{LR}M_{RR}^{-1}m_{LR}^T
\label{seesaw}
\eeq
This is the see-saw mechanism \cite{seesaw,seesaw2}. It not only generates
Majorana mass terms of the type $m_{LL}$, but also naturally makes them 
smaller than the Dirac mass terms by a factor of $m_{LR}/M_{RR}\ll 1$.
One can think of the heavy right-handed neutrinos as being integrated
out to give non-renormalisable Majorana operators suppressed
by the heavy mass scale $M_{RR}$. The goal of see-saw model
building is to choose input see-saw matrices
$m_{LR}$ and $M_{RR}$ that will give rise to Eq.\ref{hier}.

We now show how the input see-saw matrices can be simply chosen to 
give Eq.\ref{hier} with a 
naturally small sub-determinant as in Eq.\ref{det}
and hence a natural neutrino mass
hierarchy using a mechanism
first suggested in \cite{King:1998jw}.
The idea was developed in \cite{King:1999cm} where it was called
single right-handed neutrino dominance (SRHND) . SRHND was first successfully
applied to the LMA MSW solution in \cite{King:2000mb}.
The most recent discussion of these ideas is in \cite{King:2002nf}.

The SRHND mechanism is most simply described 
assuming three right-handed neutrinos
in the basis where the right-handed neutrino mass matrix is diagonal
although it can also be developed in other bases 
\cite{King:1999cm,King:2000mb}. In this basis we write the input
see-saw matrices as
\begin{equation}
M_{RR}=
\left( \begin{array}{ccc}
X' & 0 & 0    \\
0 & X & 0 \\
0 & 0 & Y
\end{array}
\right), \ \ 
m_{LR}=
\left( \begin{array}{ccc}
a' & a & d    \\
b' & b & e \\
c' & c & f
\end{array}
\right) 
\label{dirac}
\end{equation}
In \cite{King:1998jw} it was suggested that one of the right-handed neutrinos 
may dominante the contribution to $m_{LL}$ if it is lighter than
the other right-handed neutrinos. 
The dominance condition was subsequently generalised to 
include other cases where the right-handed neutrino may be
heavier than the other right-handed neutrinos but dominates due to its larger
Dirac mass couplings \cite{King:1999cm}.
In any case the dominant neutrino may be taken to be the third one 
without loss of generality. It was subsequently shown how to
account for the LMA MSW solution with a large solar angle
\cite{King:2000mb} by careful consideration of the sub-dominant contributions. 
One of the examples considered in \cite{King:2000mb} is when the 
right-handed neutrinos dominate sequentially,
\beq
\frac{|e^2|,|f^2|,|ef|}{Y}\gg
\frac{|xy|}{X} \gg
\frac{|x'y'|}{X'}
\label{srhnd}
\eeq
where $x,y\in a,b,c$ and $x',y'\in a',b',c'$.
Assuming SRHND with sequential sub-dominance as in
Eq.\ref{srhnd}, then Eqs.\ref{seesaw}, \ref{dirac} give
\beq
m_{LL}
\approx
\left( \begin{array}{ccc}
\frac{a^2}{X}+\frac{d^2}{Y}
& \frac{ab}{X}+ \frac{de}{Y}
& \frac{ac}{X}+\frac{df}{Y}    \\
.
& \frac{b^2}{X}+\frac{e^2}{Y} 
& \frac{bc}{X}+\frac{ef}{Y}    \\
.
& .
& \frac{c^2}{X}+\frac{f^2}{Y} 
\end{array}
\right)
\label{one}
\eeq
where the contribution from the first right-handed neutrino have been
neglected, but the small sub-leading contributions $xy/X$ have been included. 
If the Dirac mass couplings satisfy the condition 
$d\ll e\approx f$ \cite{King:1998jw}
then the matrix in Eq.\ref{one} resembles Eq.\ref{hier} 
and furthermore has a naturally small sub-determinant as in
Eq.\ref{det}. The neutrino mass
spectrum consists of a mass hierarchy $m_1^2\ll m_2^2\ll m_3^2$
where the heaviest neutrino has mass $m_3\approx (e^2+f^2)/Y$
\cite{King:1998jw}. The atmospheric angle is
$\tan \theta_{23} \approx e/f$ \cite{King:1998jw}.
Ignoring phases, the solar angle only depends
on the sub-dominant couplings and is given by 
$\tan \theta_{12} \approx a/(c_{23}b-s_{23}c)$ \cite{King:2000mb}.
The simple requirement for large solar angle is then $a\sim b-c$ 
\cite{King:2000mb}. Including phases the solar angle is given by 
an analagous result \cite{King:2002nf}.
A related phase analysis which pointed out the importance of phases
for determining the solar angle was given in \cite{Lavignac:2002gf}.
One also obtains the interesting bound on the angle $\theta_{13}$,
\cite{King:2002nf,Lavignac:2002gf,Akhmedov:1999uw,Feruglio:2002af}
$\theta_{13}\simgt  \frac{m_2}{m_3} \sim 0.1$.
There is therefore a good chance
that this angle could be observed at MINOS or CNGS.
The importance of charged lepton contributions to the CHOOZ
angle is explicitly discussed in \cite{King:2002nf}.

\section{A String Inspired Supersymmetric Pati-Salam Model}

The motivation for considering the SUSY Pati-Salam gauge group \cite{PaSa}
rather than conventional SUSY GUT theories is two-fold. Firstly it
avoids the infamous doublet-triplet splitting problem
\cite{Antoniadis:1988cm}, and 
secondly it may be easier to embed the gauge group directly into
a string theory \cite{Antoniadis:1990hb}.
The problem of fermion masses in such string-inspired Pati-Salam
models was first considered in \cite{King:fv}, and these ideas
were developed in detail in \cite{Allanach:1995sj}, \cite{Allanach:1995ch}.
The Pati-Salam gauge group \cite{PaSa},
supplemented by a $U(1)$ family symmetry, was first considered in
\cite{Allanach:1996hz},
\begin{equation}
SU(4)_{PS} \otimes SU(2)_L \otimes SU(2)_R\otimes U(1)_{{\rm Family}}
\end{equation}
with left (L) and right (R) handed 
fermions transforming as $F_L\sim (4,2,1)$
and $F_R\sim ({4},1,2)$ and 
the Higgs $h\sim (1,2,2)$ containing the two MSSM Higgs doublets,
\begin{equation}
{F^i_{L,R}}=
\left(\begin{array}{cccc} 
{u}  & {u} & {u} & {\nu} \\  
{d} & {d} & {d} & {e^-}     
\end{array} \right)_{L,R}^i, \ 
h=
\left(\begin{array}{cc}
{h_1}^0 & {h_2}^+ \\   
{h_1}^- & {h_2}^0      
\end{array} \right)     
\end{equation}
We assume the symmetry breaking sector of the 
minimal SUSY Pati-Salam model \cite{King:1997ia} with
further Higgs $H,\bar{H}$ transforming as 
$H\sim (4,1,2)$, $\bar{H}\sim (\bar{4},1,2)$ 
and developing VEVs which break the Pati-Salam group,
while $\theta, \bar{\theta}$ are Pati-Salam singlets
and develop VEVs which break the $U(1)$ family symmetry.
We assume for convenience that all symmetry breaking
scales are at the GUT scale.

We now consider how to describe quark and lepton (including neutrino)
masses and mixing angles, following \cite{King:2000ge}. 
The assumed $U(1)_{{\rm Family}}$ charges are
\beq
F_i=(1,0,0), \ \ F^c_i=(4,2,0)
\eeq
which leads to the universal mass matrices
\beq
m_{LR}\sim
\left( \begin{array}{ccc}
\epsilon^5 & \epsilon^3 & \epsilon    \\
\epsilon^4 & \epsilon^2 & 1 \\
\epsilon^4 & \epsilon^2 & 1
\end{array}
\right), \ 
M_{RR}\sim
\left( \begin{array}{ccc}
\epsilon^8 & \epsilon^6 & \epsilon^4    \\
\epsilon^6 & \epsilon^4 & \epsilon^2 \\
\epsilon^4 & \epsilon^2 & 1
\end{array}
\right) 
\label{m1}
\eeq
where the $\epsilon$ is some expansion parameter.
The complete fermion mass operators have the form
\begin{equation}
(F^i {F}^c_j )h\left(\frac{H\bar{H}}{M^2}\right)^n
\left(\frac{\theta}{M}\right)^p
\end{equation}
where each factor of $H\bar{H}$ corresponds to a further expansion
parameter power $\delta$, with each element of the matrix having
different Clebsch factors, giving vertical mass splittings within a generation.
The resulting set of mass matrices, suppressing the numerical
Clebsch factors and order unity coefficients, are as follows:
\beq
m^{\nu}_{LR}\sim
\left( \begin{array}{ccc}
\delta^3\epsilon^5 & \delta \epsilon^3 & \delta \epsilon    \\
\delta^3 \epsilon^4 & \delta^2 \epsilon^2 & \delta  \\
\delta^3 \epsilon^4 & \delta^2 \epsilon^2 & 1
\end{array}
\right), \ 
m^E_{LR}\sim
\left( \begin{array}{ccc}
\delta \epsilon^5 & \delta \epsilon^3 & \delta \epsilon    \\
\delta \epsilon^4 & \delta \epsilon^2 & \delta^2 \\
\delta \epsilon^4 & \delta \epsilon^2 & 1
\end{array}
\right) 
\label{m2}
\eeq
\beq
m^{U}_{LR}\sim
\left( \begin{array}{ccc}
\delta^3\epsilon^5 & \delta^2 \epsilon^3 & \delta^2 \epsilon    \\
\delta^3 \epsilon^4 & \delta^2 \epsilon^2 & \delta^3  \\
\delta^3 \epsilon^4 & \delta^2 \epsilon^2 & 1
\end{array}
\right), \ 
m^D_{LR}\sim
\left( \begin{array}{ccc}
\delta \epsilon^5 & \delta^2 \epsilon^3 & \delta^2 \epsilon    \\
\delta \epsilon^4 & \delta \epsilon^2 & \delta^2 \\
\delta \epsilon^4 & \delta \epsilon^2 & 1
\end{array}
\right) 
\label{m3}
\eeq
where $\epsilon \approx \delta \approx 0.22$.
An important role is played by ``Clebsch zeroes'' which forbid certain
leading order operators. For example Clebsch zeroes in the 
23 position forbid the leading 23 elements for the charged fermions
but allow a leading 23 element for the neutrinos with a Clebsch factor
2 (not shown explicity). 
The neutrino sector has SRHND with the third right-handed neutrino
dominating due to its large Dirac mass couplings which overcomes
the fact that the third right-handed neutrino is the heaviest one.
A large atmospheric angle is given from
the large entry in the 23 position of the ``lop-sided'' 
Dirac neutrino mass matrix 
$\tan \theta_{23} \approx e/f \sim 1$, once the Clebsch factors
and order one coefficients are included.
The solar angle is $\tan \theta_{12}\sim a/(b-c) \sim 1$.
The neutrino mass hierarchy is consistent with the LMA MSW solution.
This model has also been shown to give successful thermal leptogenesis
since the dominant right-handed neutrino is the heaviest \cite{Hirsch:2001dg}. 
It also predicts a large rate for 
$\tau \rightarrow \mu \gamma$ partly due to large $\tan \beta \sim 50$ 
also due to the large 23 element of the neutrino Dirac mass matrix
which leads large 23 contributions to the slepton
mass matrix after renormalisation group running \cite{Blazek:2001zm}.

\begin{figure}[th]
\centerline{\epsfxsize=2.5in\epsfbox{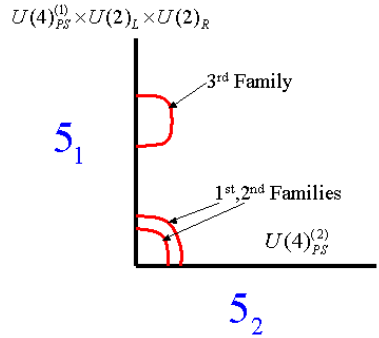}}   
\caption{A D-brane set from which the 
SUSY Pati-Salam model may emerge.\label{fig2}}
\end{figure}

The above model is not based on a unified gauge group, and therefore
apparently gauge unification cannot be implemented in such a model.
However from the point of view of string theory the gauge couplings 
may be unified along with gravity at the string scale.
As recently discussed \cite{Everett:2002pm} the
Pati-Salam model may emerge from an intersecting 
D-brane model as shown in Fig.\ref{fig2}. The assumed symmetry breaking is
\beq
U(4)_{PS}^{(1)}\times U(4)_{PS}^{(2)} \rightarrow U(4)_{PS}
\eeq
Higgs states are present which can lead to such a breaking.
The $U(1)$s are broken by the Green-Schwartz mechanism, but one
$U(1)$ remains. In fact the $U(1)$s make it hard to decouple the
exotics. In the limit that the radii of compactification satisfy
$R_2\ll R_1$ we have a ``single brane limit'' with approximate gauge
unification. The model predicts the soft mass sum rule
\beq
m_h^2+m_{F_3}^2+m_{F^c_3}^2=M_3^2
\eeq
where $m_h$ is the Higgs soft mass, 
$m_{F^c_3}$ is the right-handed third family soft mass,
$m_{F_3}$ is the left-handed third family soft mass,
and $M_3$ is the gluino soft mass, all evaluated at the high energy 
scale.

\end{document}